\documentclass[useAMS,usenatbib]{mn2e}
\usepackage{epsfig,graphics}

\def\Re{\mbox{$R_{\rm eff}$}}

\def\Msun{\mbox{$M_\odot$}}

\def\ML{\mbox{$M/L$}}
\def\Yst{\mbox{$\Upsilon_{*}$}}

\def\mst{\mbox{$M_{\star}$}}

\def\lsim{\mathrel{\rlap{\lower3.5pt\hbox{\hskip0.5pt$\sim$}}
    \raise0.5pt\hbox{$<$}}}                % less than or approx. symbol
\def\gsim{~\rlap{$>$}{\lower 1.0ex\hbox{$\sim$}}}

\def\Msun{\mbox{$M_\odot$}}

\def\lsim{\mathrel{\rlap{\lower3.5pt\hbox{\hskip0.5pt$\sim$}}
    \raise0.5pt\hbox{$<$}}}
\def\gsim{~\rlap{$>$}{\lower 1.0ex\hbox{$\sim$}}}

\def\Re{\mbox{$R_{\rm eff}$}}

\def\mst{\mbox{$M_{*}$}}

\def\gZ{\mbox{$\nabla_{\rm Z}$}}
\def\gage{\mbox{$\nabla_{\rm age}$}}
\def\gML{\mbox{$\nabla_{\rm \tiny \Yst}$}}

\def\ggr{\mbox{$\nabla_{\rm g-r}$}}

\def\ggK{\mbox{$\nabla_{\rm g-K}$}}

\def\Ngr{\mbox{$\rm N_{\rm gal}$}}

\title[Gradients and environment]{Population Gradients in the SDSS Galaxy Catalog. The role of merging}

\author[Tortora et al.]{\noindent
C.~Tortora$^{1}$\thanks{E-mail: ctortora@physik.uzh.ch},
N.R.~Napolitano$^{2}$
\\~\\
$^1$ Universit$\ddot{a}$t Z$\ddot{u}$rich, Institut f$\ddot{u}$r
Theoretische Physik, Winterthurerstrasse 190,
CH-8057, Z$\ddot{u}$rich, Switzerland\\
$^2$ INAF -- Osservatorio Astronomico di Capodimonte, Salita
Moiariello, 16, 80131 - Napoli, Italy}

\begin{document}
\date{Accepted  Received }
\pagerange{\pageref{firstpage}--\pageref{lastpage}} \pubyear{xxxx}
\maketitle

\label{firstpage}

\begin{abstract}
We investigate the role of the environment on the colour and
stellar population gradients in a local sample of $\sim 3500$
central and $\sim 1150$ satellite SDSS early-type galaxies (ETGs).
The environment is parameterized in terms of the number of
satellite galaxies, \Ngr\ in each group. For central galaxies, we
find that both optical colour and mass-to-light (\ML) ratio
gradients are shallower in central galaxies residing in denser
environments (higher \Ngr). This trend is driven by metallicity
gradients, while age gradients appear to be less dependent on the
environment and to have a larger scatter. On the other hand,
satellites do not show any differences in terms of the
environment. The same results are found if galaxies are classified
by central age, and both central and satellite galaxies have
shallower gradients if they are older and steeper gradients if
younger, satellites being independent of ages. In central
galaxies, we show that the observed trends can be explained with
the occurrence of dry mergings, which are more numerous in denser
environments and producing shallower colour gradients because of
more uniform metallicity distributions due to the mixing of
stellar populations, while no final clues about merging occurrence
can be obtained for satellites. Finally we discuss all systematics
on stellar population fitting and their impact on the final
results.
\end{abstract}


\section{Introduction}\label{sec:intro}

Colour and stellar population gradients in galaxies are providing
important clues to galaxy evolution (\citealt{Hopkins+09a};
\citealt{Spolaor+09}; \citealt{Kuntschner+10};
\citealt{Pipino+10}; \citealt{Rawle+10}; \citealt{Spolaor+10};
\citealt[herafter T+10]{Tortora+10CG}; \citealt{Tortora+10CGsim};
\citealt[herafter T+11]{Tortora+11MtoLgrad},
\citealt{LaBarbera+2011}). The value of metallicity and age
gradients and their trends with the mass have been recently
investigated on samples of local ETGs, suggesting that different
physical phenomena could concur to shape the gradients at low and
high masses (e.g. T+10). From one side, gas infall and supernovae
feedback (\citealt{Larson74, Larson75}; \citealt{Kawata01};
\citealt{KG03}; \citealt{Ko04}; \citealt{Pipino+10};
\citealt{Tortora+10CGsim}) seem to be the main phenomena driving
the evolution of low mass ETGs, while merging and AGN feedback
(\citealt{Ko04}; \citealt{Hopkins+09a}; \citealt{Sijacki+07})
would work at larger masses (\citealt{dek_birn06}). However,
environment also plays a crucial role in galaxy evolution, since
many physical phenomena like tidal interactions, strangulations
and harassment would affect the star formation in low mass
galaxies (e.g. \citealt{weinmann09}) and cause inner population
gradients at different mass scales (e.g.
\citealt{Tortora+10CGsim}). In more massive systems, though, the
major player in driving the stellar population mixing is galaxy
merging (\citealt{Davis+85}; \citealt{Springel+05};
\citealt{Romeo+08}). \cite{Ko04} have shown that overall merging
events tend to flatten the metallicity gradients with time.
Looking in more details, merging events can have a complex
taxonomy: thus, minor or major merging, or even gas-rich or -poor
merging are expected to produce different stellar population
gradients. For instance, after the initial gas rich-merging events
(generally occurring at high redshift), larger central metallicity
and positive age gradients are observed (\citealt{Ko04},
\citealt{MH94}); subsequent gas poor-merging may dilute the
positive age gradient with time as well as make the metallicity
gradients flattened out (\citealt{White80}, \citealt{Hopkins+09a},
\citealt{DiMatteo09}).

Although higher mass systems would be the ones which have
experienced a larger fraction of merging events, it is also
interesting to investigate if central galaxies in groups and
clusters may differ, in their population gradient properties, from
satellite systems orbiting in the cluster potential. The central
galaxies in clusters (and groups) are the most luminous and
massive (in terms of stellar mass) objects in the Universe. They
are found to have different luminosity profiles when compared with
typical cluster galaxies (\citealt{Schombert86}) and do not seem
to be drawn from the same luminosity function as bright
ellipticals (\citealt{Dressler78}, \citealt{BB01}). These
evidences suggest that the evolutive processes of central galaxies
can be strongly different from the ones driving normal (satellite)
systems. Moreover, they still hold imprints from the early
evolutionary stages since they reside in the very central regions
of clusters and groups, where mass started to be accreted earlier
after the Big Bang than other density environments. These regions
have witnessed during the cosmic time a variety of galaxy
interactions with the environment, and merging events (e.g.
\citealt{Romeo+08}) contributing to the mass accretion of larger
and larger galaxy systems, a process which is more efficient in
denser environments like group/cluster haloes (e.g.
\citealt{Whiley+08}, \citealt{Stott+08}, \citealt{Stott+10}).

From this perspective it is natural to expect that the galaxy in
the center of very dense environments might be more sensitive to
the effect of the large amount of merging events expected in the
hierarchical growth which shall be recorded in the stellar
population parameters (e.g. age and metallicity).

Thus, we have considered a local sample of SDSS galaxies
(\citealt{Blanton05}), recently analyzed and discussed in T+10 and
T+11, where we have discussed colour, mass-to-light ratio and stellar
population gradients in terms of mass and compared with
independent observations. These results have been framed within
the predictions of hydrodynamical and chemo-dynamical simulations
of galaxy formation. In the present paper we will discuss the
connection with the environment, selecting those
galaxies classified as centrals and satellites in groups and
clusters and investigating if gradients change as a function
of the environment.

The paper is organized as follows. The data sample and the
analysis have been presented in \S \ref{sec:sample}, the results
are discussed in \S \ref{sec:results}, while \S
\ref{sec:conclusions} is devoted to the physical interpretation
and conclusions. Systematics in stellar population fit have been
discussed in \S \ref{sec:app_syst}.

\begin{figure*}
\psfig{file= 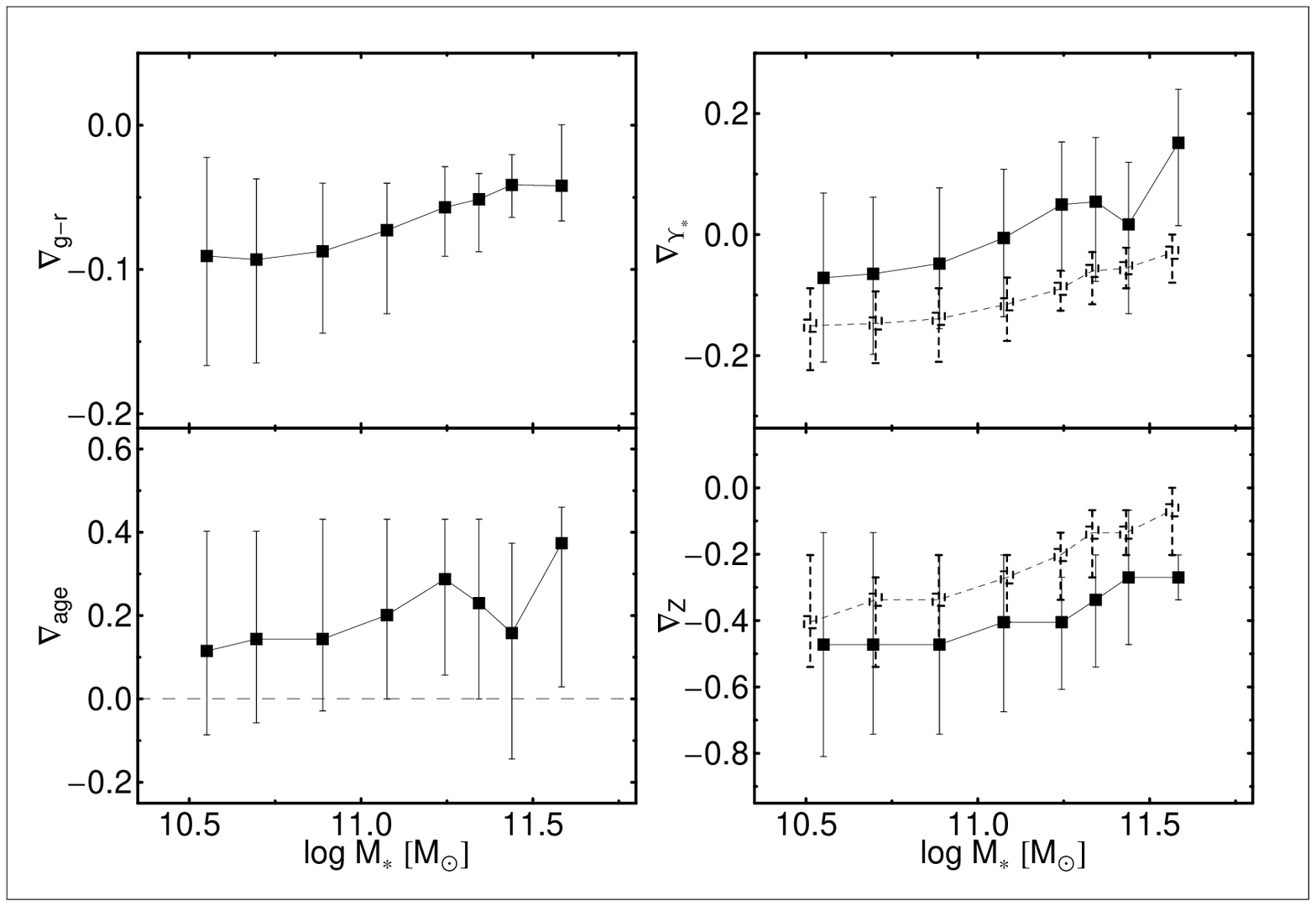, width=0.45\textwidth} \hspace{0.5cm}
\psfig{file= 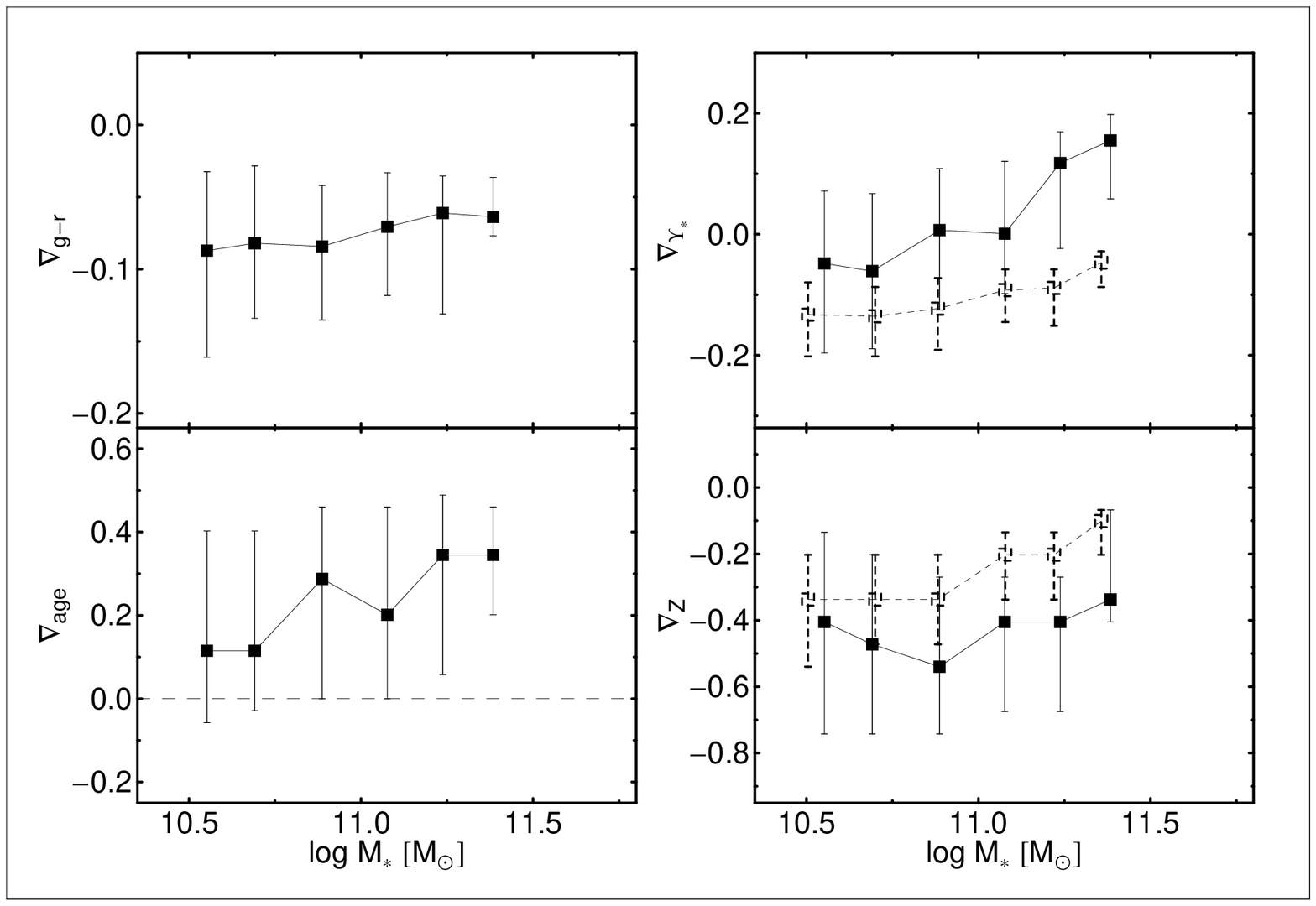, width=0.45\textwidth} \caption{Gradients
in terms of stellar mass for centrals (left) and satellites
(right). In each panel we show $g-r$ (top left), \ML\ (top right),
age (bottom left) and Z (bottom right) gradients as a function of
stellar mass. The medians and 25-75th quantiles are shown. The
dashed symbols are for the case with $\gage=0$ (i.e. $\rm age_{1}=
age_{2}= 10 \, \rm Gyr$).}\label{fig: fig1}
\end{figure*}

\section{Data and spectral analysis}\label{sec:sample}

We start from a database consisting of $50\,000$ low redshift
($0.0033 \leq z \leq 0.05$) galaxies in the NYU Value-Added Galaxy
Catalog extracted from SDSS DR4 (\citealt[hereafter
B05]{Blanton05})\footnote{The catalog is available at: {\tt
http://sdss.physics.nyu.edu/vagc/lowz.html}.}, recently analyzed
in T+10 and T+11\footnote{Details about sample selection,
incompleteness and biases can be found in T+10.}, where stellar
population synthesis models have been used to determine galaxy
stellar mass, \mst, colour and stellar population
parameters/gradients. Following T+10, we have sorted out ETGs by
keeping those systems with a S$\rm \acute{e}$rsic index satisfying
the condition $2.5 \leq n \leq 5.5$ and with a concentration index
$C
> 2.6$. The final ETG sample consists of $10\,508$ galaxies. We
have cross-matched our datasample with the ($z>0.01$) DR4 SDSS
based group catalog from \cite{Yang+07} to recover information
about the environment the galaxies live. \cite{Yang+07} have
identified the groups and have separated the most massive (or most
luminous) galaxies in each group, labeled as centrals, from the
satellites. We will use both 1) the identification of satellite
and centrals on the base of stellar mass selection, and 2) the
number of galaxies $N_{\rm gal}$ in each group as an environment
indicator. We have retained galaxies -- isolated (having $N_{\rm
gal} = 1$), central and satellite -- with a mass $\log \mst >
10.5$ and left with a final sample of 3525 central (including 1941
isolated systems) and 1141 satellite galaxies. As a further
criterion, we have also ranked satellites in each group on the
basis of their stellar masses.

We notice that the fraction of central ETGs in each mass bin is
lower than the fraction of satellite and field galaxies at $\log
\mst/\Msun \lsim 10.7-10.9$, while it increases at larger masses,
with 62 centrals, 6 satellites and 1 field galaxy for $\log
\mst/\Msun > 11.4$.

As discussed in T+10, we have used the structural parameters given
by B05 to derive the color profile $(X-Y)(R)$ of each galaxy as
the differences between the (logarithmic) surface brightness
measurements in the two bands, $X$ and $Y$. The stellar population
properties are derived by the fitting of \citet[hereafter
BC03]{BC03} ``single burst'' synthetic stellar models to the
observed colours. Age and metallicity are free to vary, and a
\cite{Chabrier01} IMF is assumed. However, in order to check the
effect of the existing degeneracies between age and metallicity,
we will also assume the age gradient to zero
(\citealt{Tortora+11MtoLgrad}) by fixing the age to $10 \, \rm
Gyr$ as a prior\footnote{{As we will also discuss later, $\gage =
0$ is a reasonable assumption for the older galaxies, which are
found to have $\gage \sim 0$ (\citealt{Tortora+10CG}).}}. We
define the CG as the angular coefficient of the relation $X-Y$ vs
$\log R/\Re$, $\displaystyle \nabla_{X-Y} = \frac{\delta
(X-Y)}{\delta \log (R/R_{\rm eff})}$, measured in $\rm mag/dex$
(omitted in the following), where \Re\ is the r-band effective
radius. By definition, a positive CG, $\nabla_{X-Y}>0$, means that
a galaxy is redder as $R$ increases, while it is bluer outward for
a negative gradient. The fit of synthetic colours is performed on
the colours at $R_{1} = \Re / 10$ and $R_{2} = \Re$ and on the
total integrated colours. Stellar parameter gradients are defined
as $\displaystyle \nabla_{W} = \frac{\delta \log (W)}{\delta \log
(R/R_{\rm eff})}$, where $W = (age, Z, \Yst)$ are the estimated
age, metallicity and \ML. Because of the definitions adopted for
$R_{1}$ and $R_{2}$, an easier and equivalent definition can be
used, in fact, we define $\gage =\log [\rm age_{2}/\rm age_{1}]$,
$\gZ=\log [Z_{2}/ Z_{1}]$ and $\gML=\log [\rm \Yst_{2}/ \rm
\Yst_{1}]$ where (age$_{i}$, $Z_{i}$, $\Yst_{i}$), with $i=1,2$,
are the estimated parameters at $R_{1}$ and $R_{2}$, respectively.

In T+10 and T+11 we have discussed the impact of the wavelength
coverage on the results. Here, we will add further Montercarlo
simulations in order to verify that the optical bands alone
introduce small spurious degeneracies and/or correlations among
parameters with respect to stellar models including near-IR or UV
data. In particular, we will quantify any systematics and possible
trends with stellar mass and \Ngr\ in the Appendix \ref{fig:
fig_app_1}, and discuss the impact on our results in \S
\ref{sec:conclusions}.

\section{Gradients and environment}\label{sec:results}

In T+10 and T+11 we have discussed the colour, age, Z and \ML\
gradients as a function of stellar mass without distinguishing
central from satellite galaxies. Here, we will expand the analysis
by investigating the effect of the environment on the most massive
central and satellite galaxies, showing the significance of the
correlations in Table \ref{tab:tab1} and the slopes in Table
\ref{tab:tab2}.

We start in Fig. \ref{fig: fig1} by showing the gradients \ggr,
\gML, \gZ\ and \gage\ as a function of the stellar mass for
central and satellite galaxies. The trend of central galaxies
alone does not differ much from the one of the total sample
studied in T+10 and T+11. They have colour and metallicity
gradients which become shallower at larger masses. \gML s are
negative at low masses, turn out to be null around $\log
\mst/\Msun \sim 11$ and finally become slightly positive at very
high masses. The positive trend of \gage\ is statistically
significant (as shown in Tables \ref{tab:tab1} and \ref{tab:tab2})
despite its large scatter. \gZ\ show an increasing trend with
\mst\ (from -0.5 to -0.3 across the mass range), although the
scatter at lower masses is quite large.

If one forces $\gage = 0$ (assuming $age_{1}= age_{2}= 10 \, \rm
Gyr$), \gZ s are rigidly shifted toward larger values and show a
steeper trend with stellar mass, with $\gZ \sim 0$ for the most
massive galaxies. Similarly, \gML\ are shifted toward more
negative values while their trend with mass is shallower (see
T+11). Note that in case of no age gradients, (negative) shallower
\gML s naturally correspond to (negative) shallower colour and
metallicity gradients. On the contrary, in presence of non--zero
age gradients negative colour gradients could correspond to
positive \ML\ gradients due to recent star formation episodes.
Almost all of these correlations are significant at more than
$99\%$ (see Table \ref{tab:tab1}).

Satellite galaxies behave differently from centrals. Colour
gradients are less dependent on mass, while the \gML\ trend is
steeper with mass. This is driven by the \gage\ which is
increasing at the higher mass bins, being \gZ\ less dependent on
\mst. The fact that the \gML\ trend is mainly driven by the age
gradients is confirmed when forcing $\gage = 0$: in this case the
trend of \gML\ with mass gets shallower although \gZ\ show a steep
increasing trend with \mst.

\begin{figure*}
\psfig{file= 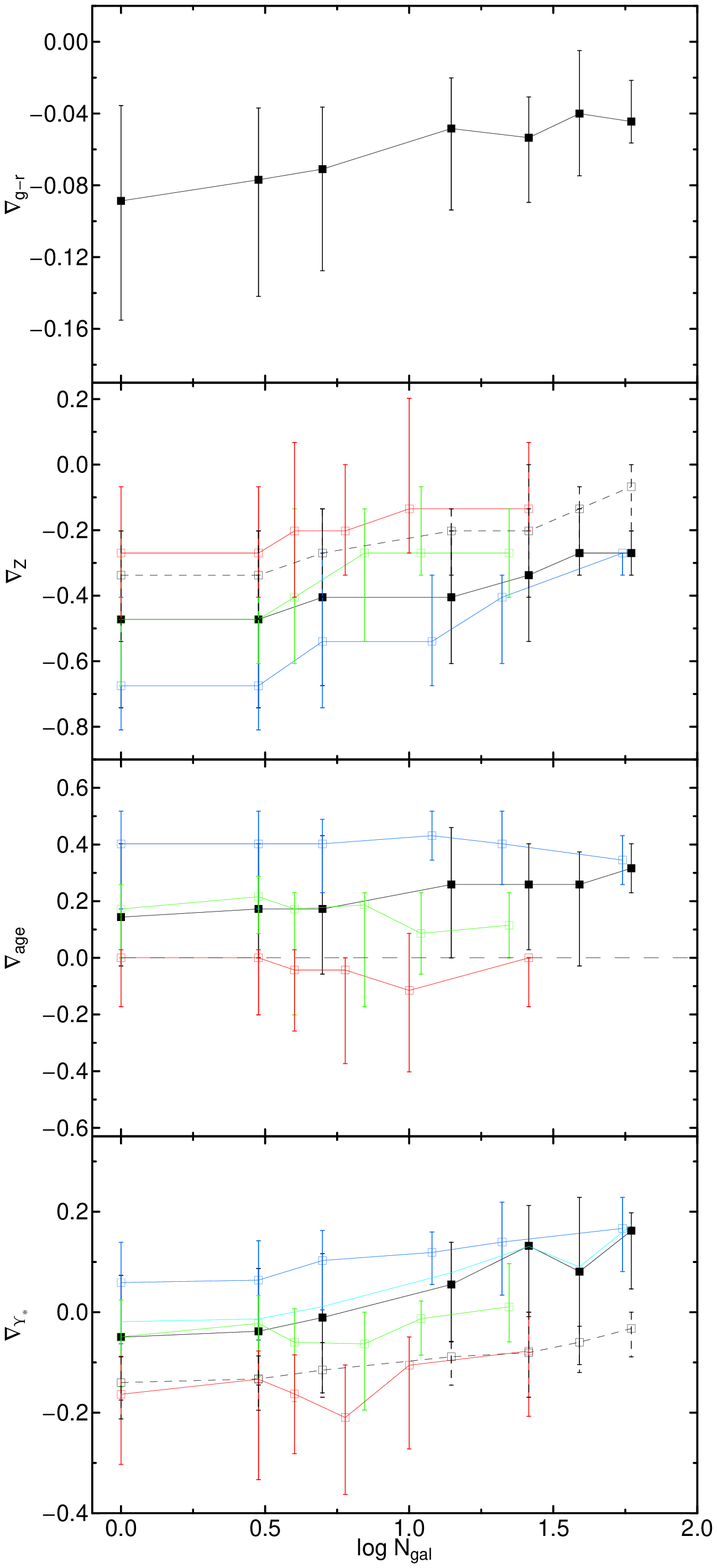, width=0.45\textwidth}
\hspace{0.3cm}\psfig{file= 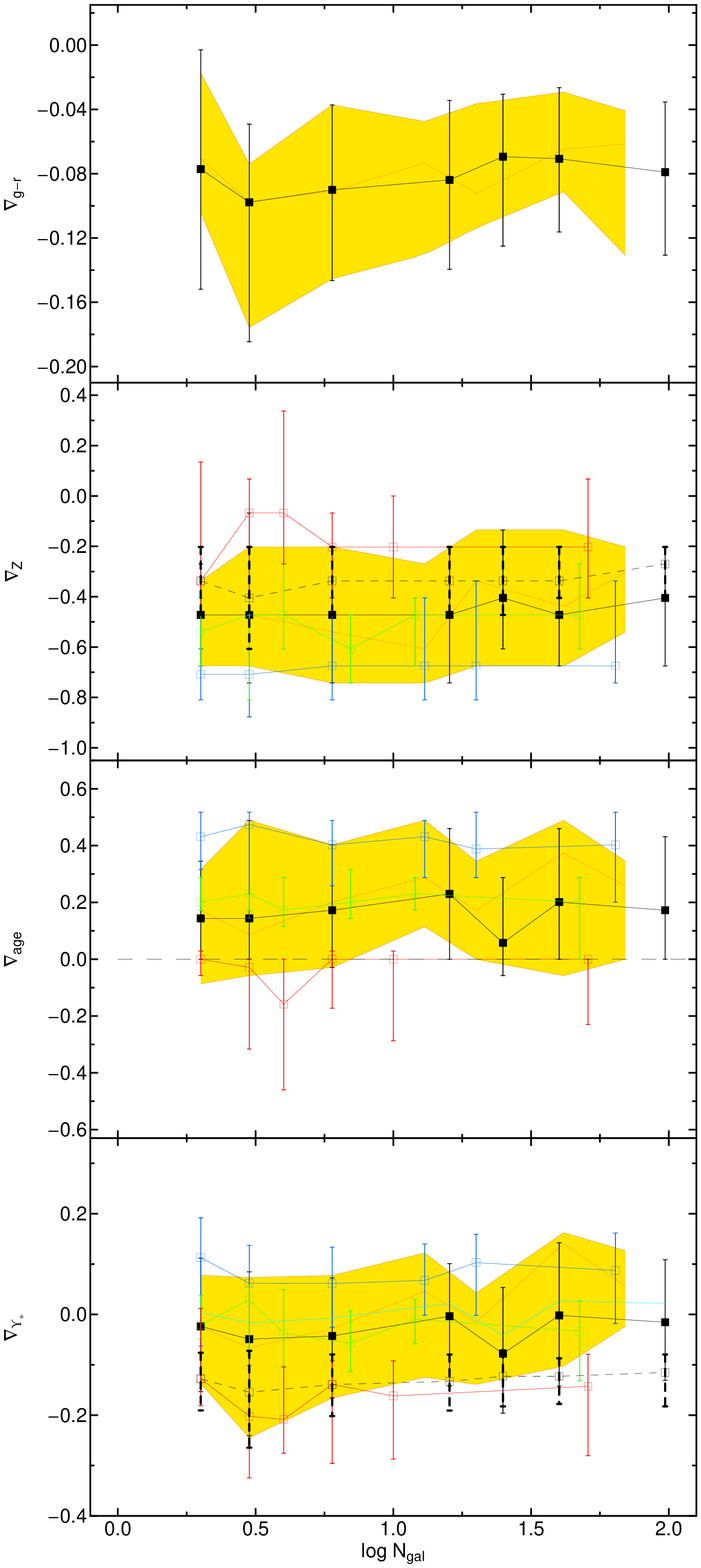, width=0.45\textwidth}
\caption{Gradients in terms of the environment for centrals (left)
and satellites (right). Medians and 25-75th quantiles are shown.
From the top to the bottom, colour, metallicity, age and \ML\
gradients are plotted as a function of \Ngr. The black symbols are
as in Fig. \ref{fig: fig1}. Moreover, blue, green and red symbols
are for galaxies with central age in the intervals  $0 < age_{1}
\leq 6$, $6 < age_{1} \leq 10$ and $ age_{1} > 10$ Gyr,
respectively. In the bottom panel, the cyan line is relative to
the V-band \ML\ gradients. Orange lines and yellow regions set
medians and 25-75th quantiles for the second most massive galaxy
in each group.}\label{fig: fig2}
\end{figure*}

In Fig. \ref{fig: fig2}, colour, metallicity, age and \ML\
gradients are plotted as a function of \Ngr. We  find that in
denser environments (higher \Ngr) central galaxies have shallower
colour gradients, going from $\ggr \sim -0.09$ in the field to
$\ggr \sim -0.04$ in the clusters ($\Ngr \gsim 50$). This trend is
reproduced by a similar behaviour of metallicity gradients which
are on average $\sim -0.5$ in the field and $\sim -0.3$ in denser
environments. On the contrary, \gage\ are only slightly steeper in
densest environments. According to the trends of colour and Z
gradients, \gML\ tend to be steeper in the field ($\sim -0.05$),
almost null in poor groups ($\Ngr \sim 10$), while they are
positive ($\sim 0.1$) at $\Ngr \gsim 50$. The trends with
environment are clearer if we divide the sample in terms of the
central age, $age_{1}$. Three age intervals are adopted, i.e. $0 <
age_{1} \leq 6$, $6 < age_{1} \leq 10$ and $ age_{1} > 10$, which
gather 1808, 594, 1123 centrals and 577, 198, 366 satellites,
respectively. Older galaxies have progressively shallower age and
metallicity gradients, independently from \Ngr\ (see also the
discussion in T+10). The trends with \Ngr\ are still statistically
significant for \gZ, which become shallower going from the field
to denser environments by $\sim 0.15, 0.2, 0.4$, while the
correlations for \gage\ get flat for each age bin. The behaviour
of \gML\ is different: oldest galaxies have negative \gML\ ($\sim
-0.2,-0.1$) with a shallower trend with \Ngr\ (in the very rich
systems they are almost zero), while the youngest ones have
positive gradients ($\sim 0.05, 0.15$), which get larger in higher
density environment.

On the contrary, satellite galaxies have gradients which are
completely independent of the environment they live (see slopes in
Table \ref{tab:tab2}). To have a complete view of behaviours in
satellites, we have also considered the rank-2 satellites, i.e.
the second most massive galaxy in each group and plotted their
gradients in the right panel of Fig. \ref{fig: fig2}. Here, we see
little changes in the \ggr\ and \gZ\ trends, while \gage\ and
\gML\ seem to be steeper functions of \Ngr, with steeper values in
very high density environment. To avoid a too low statistics, we
have not investigated higher rank systems.

As a comparison, we also show in Fig.~\ref{fig: fig2} the results
when galaxy colours are fitted to a synthetic spectral model with
no age gradient. In this case, metallicity gradients turn out to
be shallower, independently of \Ngr, while \ML\ gradients are
steeper. Note that these last trends are consistent with what was
found for the older galaxies in the reference fit. In centrals,
the trend is almost unchanged for \gZ, while a shallower slope il
found for \gML. For satellites, slopes resulted unchanged. For
completeness, in the bottom panel we also compare with V-band \ML\
gradients, which despite a slight shift are very similar to the
reference B-band ones.

We have finally checked that the different trends in Fig.
\ref{fig: fig2} between centrals and satellites survive if we cut
the most massive galaxies with $\log \mst/\Msun > 11.4$ or if we
include in our sample the galaxies with $10< \log \mst/\Msun \leq
10.5$.

If we bin the galaxies in terms of stellar mass we find that if
considering the lower mass galaxies the trends in Fig. \ref{fig:
fig2} for centrals disappear, while for more massive systems the
trend become more significant and also the $\Ngr -\mst$ relation
is steeper. On the contrary, if we bin the gradients-\mst\
relations in Fig. \ref{fig: fig1} in terms of \Ngr\ we find that
field galaxies show some trends in \gage\ and \gZ, which are
stronger in group centrals.

\begin{table}
\centering \caption{Sign of the correlation between gradients and
\mst\ or \Ngr\ and confidence limits. The significance of the
correlation is obtained by applying the Student's t-distribution
to the Spearman rank factor.}\label{tab:tab1} \scriptsize
\begin{tabular}{ccccc} \hline
\rm  & \multicolumn{2}{c}{Centrals} & \multicolumn{2}{c}{Satellites}  \\
\hline
$\ggr-\mst$ & \multicolumn{2}{c}{$\uparrow \, (99\%)$}  & \multicolumn{2}{c}{$\uparrow \, (95\%)$}\\
$\ggr-\Ngr$ & \multicolumn{2}{c}{$\uparrow \, (99\%)$}  & \multicolumn{2}{c}{$\uparrow \, (95\%)$}\\
  \hline
   &  All free & $\gage = 0$ &  All free & $\gage = 0$ \\
 \hline
 $\gage-\mst$ & $\uparrow \, (99\%)$  & - & $\uparrow \, (99\%)$  & -\\
 $\gZ-\mst$ & $\uparrow \, (95\%)$  & $\uparrow \, (99\%)$ & Null $(99\%)$ & $\uparrow \, (99\%)$\\
 $\gML-\mst$ & $\uparrow \, (99\%)$  & $\uparrow \, (99\%)$ & $\uparrow \, (99\%)$ & $\uparrow \, (99\%)$\\
 &   &  &  & \\
 $\gage-\Ngr$ & $\uparrow \, (90\%)$  & - & Null $(99\%)$  & -\\
 $\gZ-\Ngr$ & $\uparrow \, (99\%)$  & $\uparrow \, (99\%)$ & $\uparrow \, (95\%)$ & $\uparrow \, (99\%)$ \\
 $\gML-\Ngr$ & $\uparrow \, (99\%)$  & $\uparrow \, (99\%)$ & $\uparrow \, (95\%)$ & $\uparrow \, (95\%)$\\
 \hline
\end{tabular}
\end{table}

\begin{table}
\centering \caption{Slopes of the correlation between gradients
and \mst\ or \Ngr\ and $1 \sigma$ uncertainties derived by
bootstrap method.}\label{tab:tab2} \scriptsize
\begin{tabular}{ccccc} \hline
\rm  & \multicolumn{2}{c}{Centrals} & \multicolumn{2}{c}{Satellites}  \\
\hline
$\ggr-\mst$ & \multicolumn{2}{c}{$0.06 \pm 0.01$}  & \multicolumn{2}{c}{$0.04 \pm 0.02$}\\
$\ggr-\Ngr$ & \multicolumn{2}{c}{$0.03 \pm 0.01$}  & \multicolumn{2}{c}{$0.01 \pm 0.01$}\\
  \hline
   &  All free & $\gage = 0$ &  All free & $\gage = 0$ \\
 \hline
 $\gage-\mst$ & $0.21\pm 0.04$  & - & $0.28 \pm 0.07$  & -\\
 $\gZ-\mst$ & $0.23\pm 0.04$  & $0.30 \pm 0.04$ & $0.17\pm0.09$ & $0.26 \pm0.05$\\
 $\gML-\mst$ & $0.20 \pm 0.02$  & $0.12 \pm 0.01$ & $0.26\pm 0.06$ & $0.10 \pm 0.02$\\
 &   &  &  & \\
 $\gage-\Ngr$ & $0.08 \pm 0.04$  & - & $0.02 \pm 0.04$  & -\\
 $\gZ-\Ngr$ & $0.12 \pm 0.02$  & $0.13 \pm 0.03$ & $0.05 \pm 0.04$ & $0.05 \pm 0.02$ \\
 $\gML-\Ngr$ & $0.11 \pm 0.02$  & $0.05 \pm 0.01$ & $0 \pm 0.02$ & $0.01 \pm 0.01$\\
 \hline
\end{tabular}
\end{table}

\section{Discussion and conclusions}\label{sec:conclusions}

In this paper we have investigated the correlation between the
colour, \ML\ and stellar population gradients with environment for
a sample of local central and satellite SDSS central galaxies
(\citealt{Blanton05}). We have found some indications which point
to a different behaviour in terms of stellar mass and the
environment for central and satellite galaxies. Central galaxies
dominate the galaxy sample in the massive side, thus have
gradients which behave like in T+10 and T+11 when the gradients
are plotted in terms of \mst, with \ggr, \gML\ and \gZ\ getting
shallower and \gage\ (positive) steeper in the most massive
galaxies. On the contrary, satellite galaxies have a \ggr\ which
is less dependent on \mst. Thus, while \gage\ are steeper at
larger masses (where they drive the trend of \gML), metallicity
gradients show a shallower trend. The steeper and stronger trend
of \gZ\ and the slightly shallower \gZ\ in the most massive
centrals, when compared with satellites, are consistent with the
role of dry mergings (\citealt{Ko04}), which are more efficient to
mix stellar populations at the largest masses where their fraction
is larger (i.e. \citealt{deLucia06}). Shallower trends are found
in satellites, suggesting a minor role of such kind of phenomena.
In satellite we have found a strong trend of \gage\ with mass
(stronger that the one found in centrals), with very massive
satellites having steeper \gage, and thus younger cores. These
surviving young cores could be due to bursts of star formation
after gas-rich mergers or close encounters. In T+10 we have also
seen that central age has an important role, since centrally
younger galaxies have steeper gradients, while older systems have,
on average, null \gage\ and shallower \gZ. This seems consistent
with the intervention of processes like dry-mergings, which are
responsible for a star formation suppression and a mixing of
stellar population\footnote{A comment is in order here.
Metallicity gradients are less sensitive to recent starburst
events, while they are more indicative of the total integrated
star formation history. This because the metal content responds
more slowly to the star formation as metals are not strongly
altered by new stars, but instead by the integral of all the star
formation history. On the other hand age gradients are by
definition sensitive to (significant) star burst events. }.

Plotting the gradients in terms of the number of galaxies inside
the group, \Ngr, for central galaxies we find that age gradients
seem less dependent on the environment, while colour, \ML\ and
metallicity gradients are shallower in denser environment (with a
number of galaxies in the groups of $\Ngr \sim 50$ or more) when
compared with isolated galaxies. This correlation is even stronger
when galaxies are classified in terms of their central ages, which
have been shown to drive the scatter of the relation between
metallicity/age gradients and mass. On the contrary, massive
satellite galaxies do not  present any trend with \Ngr\ (and
independently on the age bin adopted). Consistently with the
results found in terms of stellar mass, while dry-mergings seem to
be important process in massive centrals, no strong indication is
found for satellites. Finally, we have also checked that ranking
satellites on the basis of their masses, and considering the
second most massive galaxy in each group, some slight positive
correlations seem to appear in our trends in Fig. \ref{fig: fig2}.
These findings suggest that if on one hand conclusions on centrals
are firm and clear, on the other hand satellites results are more
complicated to interpret. More detailed analysis would investigate
the correlation of population gradients not only with \Ngr, but
also with the local and global density, the distance from the
center of the groups and the mass ranking in the group, which all
together give a more complete representation of the environmental
properties.

The positive correlation found between colour and Z gradients with
both \mst\ and \Ngr\ in central galaxies are not completely
independent each others, since it is easy to show that more
massive central galaxies are residing in denser environments with
larger cluster/group haloes and populated by a larger number of
galaxies (e.g. \citealt{Whiley+08}, \citealt{Stott+08},
\citealt{Stott+10}).

The systematics on the gradients have been investigated in the
Appendix \ref{sec:app_syst}, where we have shown that \gML\ is
very little affected (see, e.g., T+11), while a shift and a larger
scatter could be induced in the estimates of \gage\ and \gZ. The
trend with \mst\ and \Ngr\ could be possibly affected, in the
sense that the true trends for \gZ\ can be shallower or flat and
the one for \gage\ steeper (in the worst cases when input data are
strongly unaccurate).

From the observational point of view, these are the first steps in
the systematic study of the correlations between stellar
population gradients and the environment as measured by \Ngr\
where the behavior of the central galaxies is distinguished by the
one of satellites (see e.g. \citealt{LaBarbera+2011}). Former
studies made no selection on the galaxy type (centrals or
satellites), and found that color gradients in massive galaxies
are shallower in denser environment (\citealt{KI05} and
\citealt{LaBarbera2005}) similarly to our findings, although,
\citealt{LaBarbera+2011} have shown that the optical-IR \ggK\ is
almost independent of the environment. Along the same line,
\cite{Roche+10} have demonstrated that central galaxies have
shallower gradients than normal E/S0 galaxies at the same
luminosity.

For what concerns stellar population gradients, \cite{Spolaor+09}
(their right panel in Figure 1) have compared central galaxies in
both clusters and groups with those in the field, but a clear
trend is not found since the scatter of the data is too large and
the sample too small. Using a combined sample of satellite and
central galaxies, \cite{LaBarbera+2011} have found that \gage\ are
slightly increasing with \mst\ (similarly to our Fig. \ref{fig:
fig1}), while \gZ\ decreases, and galaxies in groups present
steeper \gage\ and \gZ, which are in contrast with our results.
The latter discrepant results may be a consequence of a different
sample selection or systematics in our stellar population models.
Differences in the galaxy sample selection would contribute to
some of the discrepancies discussed above. In fact, they are
adopting a sample without any selection in centrals and
satellites, with the latter dominating in number, while in our
case the number of centrals is larger and if we derive the plots
in Fig. \ref{fig: fig2} for the whole sample, then no trend is
found and a better agreement is found. Then, if we consider the
worst case analyzed in simulations discussed in the Appendix (see
e.g. Fig. \ref{fig: fig_app_1}) there seems to be some spurious
trends introduced by the use of the optical bands only: if we
correct our results for this systematics we can also conclude that
a steeper trend with mass for \gage\ and a shallower trend (if
any) for \gZ\ would be accommodated by our data, in a way more
consistent with \cite{LaBarbera+2011} findings. If this is the
case, it might indicate a major role of gas-rich mergings, cold
accretion at high redshift, later gas accretion or close
encounters, in satellites, in order to produce steeper gradients
in very massive galaxies, rather than dry merging as we would
conclude if the gradients are shallower like optical data suggest.
Model systematics seem to weakly affect the conclusions with \Ngr\
though, which clearly indicates a difference of the gradient
behaviour between central galaxies and satellites.

Our results give an indication of the role of physical processes
in the mass accretion of massive central and satellite galaxies.
The effect of merging and, in general, of interactions with
environment crucially depends on the kind of process under
analysis (minor vs major merging, dry vs wet mergings, accretion,
stripping, etc. etc.). E.g., the positive age gradients found
mainly in both (centrally young) central and satellite galaxies
supports two kinds of processes: a) a dissipative formation
picture, whereby wet mergers fuel the central region with cold gas
(\citealt{deLucia06}), or b) cold accretion at high redshift
(\citealt{dek_birn06}), which generate the observed younger
stellar populations in the center. But, central galaxies have more
chances to merge or interact with neighbors in denser
environments, in fact they are predicted to have larger masses in
clusters when compared with similar galaxies in the centers of
groups (\citealt{Romeo+08}). The flatter gradients of centrals we
observe in denser environments are consistent with the
intervention of a larger number of dry (major) galaxy mergings,
which are predicted to mesh stellar populations inside the
galaxies (e.g. \citealt{White80}; \citealt{Ko04}). On the
contrary, satellite galaxies do not show any strong evidence in
favour of this kind of merging events. The observed younger cores
in the youngest galaxies suggest that mergings was not so
effective to erase these pre-existing age gradients (possibly
generated after a gas-rich merging or cold accretion) or some
recent accretion in the center has happened. While, dry mergings
were really efficient to dilute those gradients in older galaxies.

The results discussed in this paper represent an useful reference
exercise for further analysis performed on higher--z samples,
typically involving optical rest frame bandpasses (e.g. from the
VLT Survey Telescope, VST). Future analysis with wider wavelength
baselines will help to alleviate the problem of parameter
degeneracies in stellar modelling (e.g.,
\citealt{LaBarbera+2011}).

\section*{Acknowledgments}

CT was supported by the Swiss National Science Foundation.

\appendix

\section{Systematics in the stellar fit}\label{sec:app_syst}

\begin{figure*}
\psfig{file= 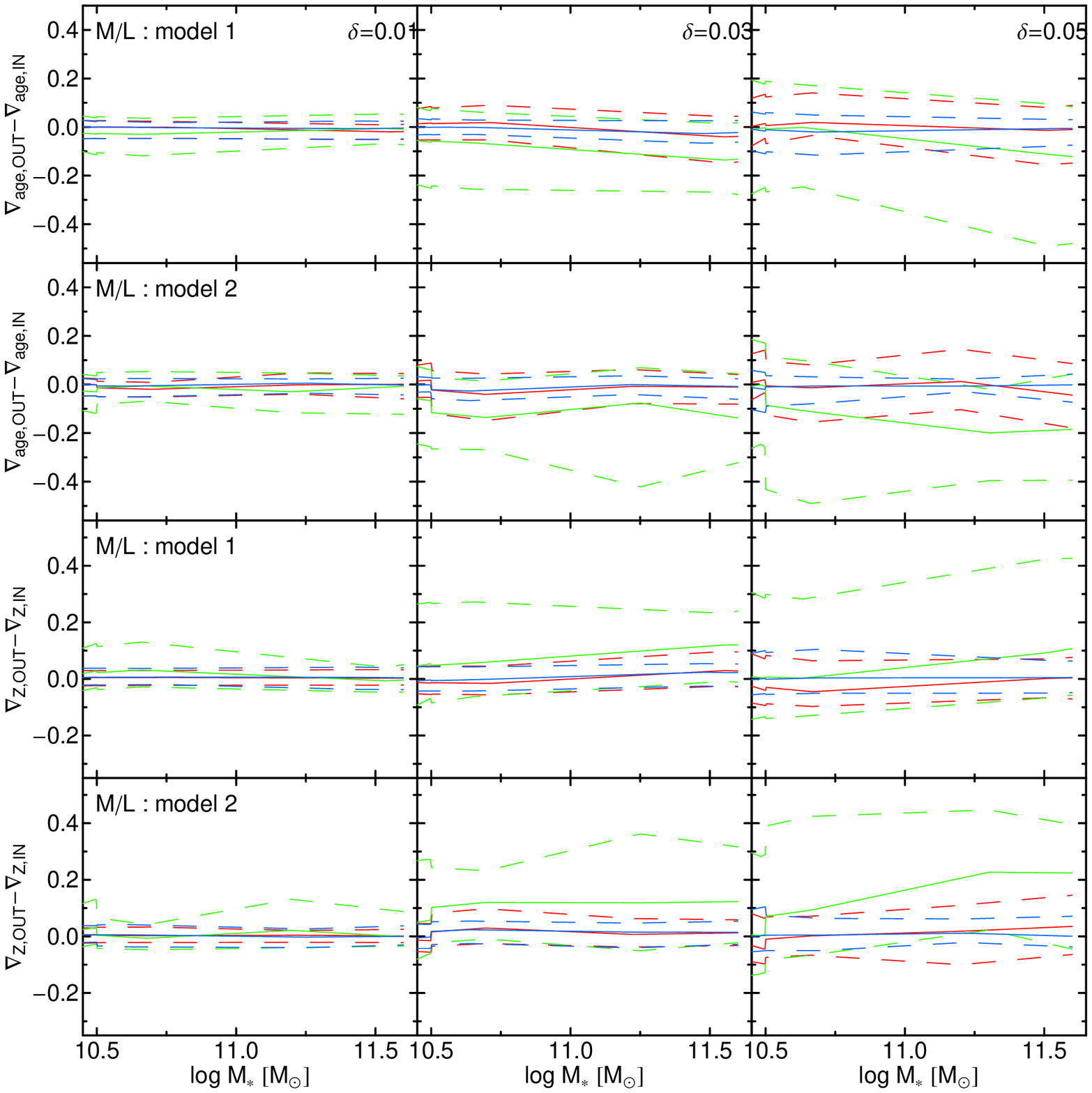, width=0.5\textwidth}\psfig{file=
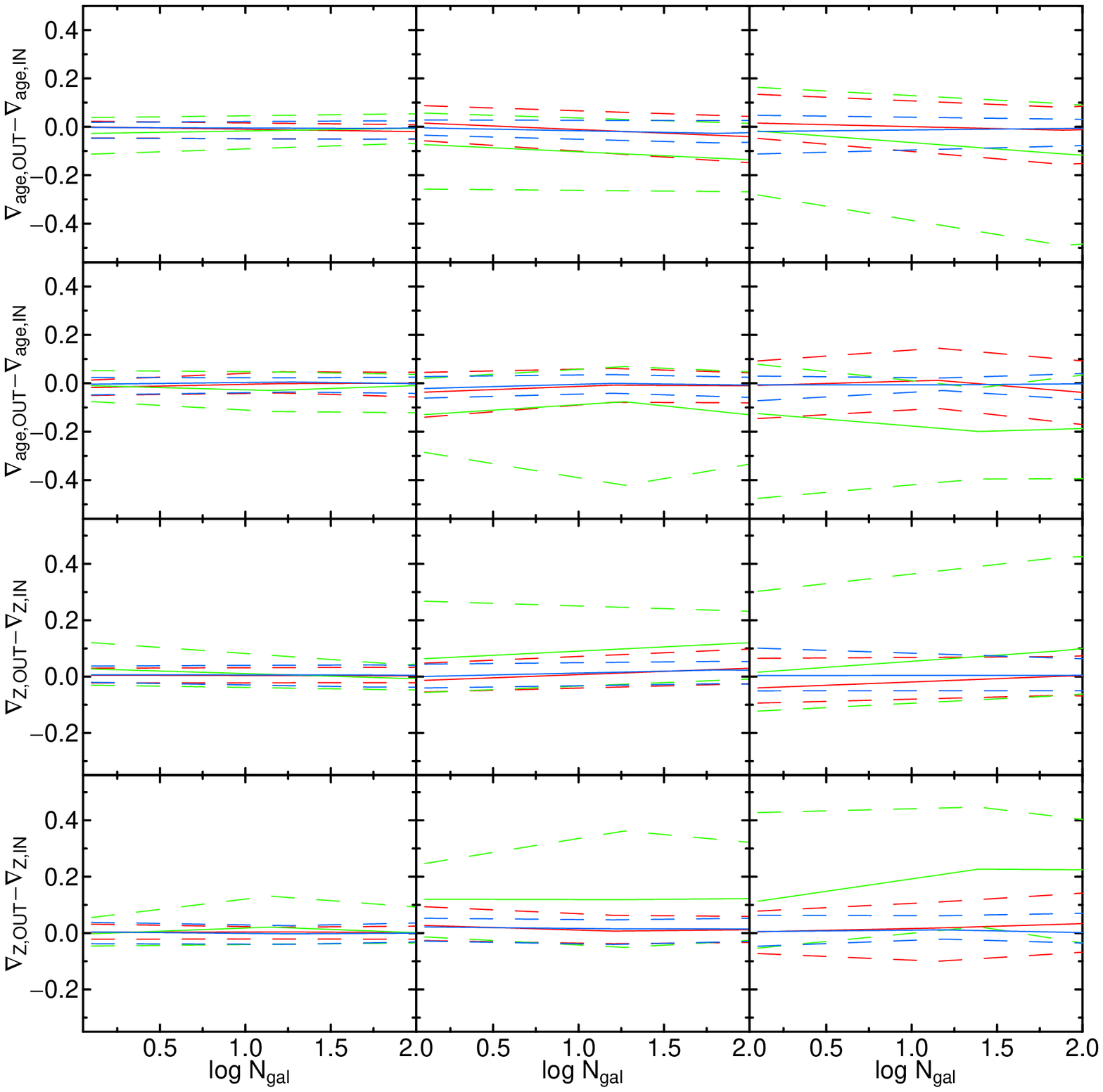, width=0.5\textwidth} \caption{Systematics in
stellar population fitting in terms of stellar mass (left panels)
and galaxy number in the groups (right panels). We define the
quantity $\nabla_{X,OUT}-\nabla_{X,IN}$, where $X= (age, Z)$, IN
and OUT are the input and output values. In each big panel, from
the top to the bottom we show the systematics in age gradients for
two \ML-\mst\ relations and Z gradients for the same two \ML-\mst\
relations. From the left to the right we show the cases $\delta =
0.01$, $0.03$ and $0.05$. The continue line is the median while
dashed ones are the 25-75th quantiles. Green, red and blue are for
the fit using optical, optical+NIR and optical+UV,
respectively.}\label{fig: fig_app_1}
\end{figure*}

Here, we check for systematic effects on our gradient (\gage\ and
\gZ) estimates from the stellar population fits. Because of the
well known age-metallicity degeneracy (\citealt{Worthey94},
\citealt{BC03}, \citealt{Gallazzi05}), the stellar population
parameters and the gradients might be biased when using the
optical colours only, as we do now. Widening the wavelength
baseline to include near-infrared (NIR) colours should ameliorate
the degeneracy. Although the synthetic prescriptions in the
near-IR region of the spectra are still uncertain, it is very
important to understand their impact on our analysis. In T+10 and
\cite{Tortora+11MtoLgrad} we checked age and metallicity
inferences using optical versus optical+NIR constraints, and
found, on average, little systematic difference. Anyway, depending
case by case, a spurious shift can appear in the estimated
gradients and a wide scatter when the fit is made only using
optical data. Here we will carry out a similar analysis to study
the trends with stellar mass and \Ngr, using a suite of Monte
Carlo simulations. Because of the large amount of details
involved, we will consider the results from these simulations as a
qualitative guide to understand in what direction the correlations
found would change.

We  extract 1000 simulated galaxy spectra from our BC03 spectral
energy distribution libraries with random, uncorrelated stellar
parameters and gradients. In order to have gradients similar to
the ones we observe, we only impose some constraints on the input
stellar parameters, producing, on average, $\gage \sim 0.2$ and
$\gZ \sim -0.3$. We add simulated (equal) measurement errors for
each band, as randomly extracted steps from the interval ($-
\delta$,$+ \delta$), with $\delta = 0.01, 0.03, 0.05$. We apply
our fitting procedure and then compare the output parameter
estimates to the input model values. We perform the fit (1) using
only the optical SDSS bands $ugriz$ and (2) adding NIR photometry
($J$, $H$ and $Ks$; \citealt{Jarrett+03}) and (3) ultraviolet
photometry (NUV and FUV; \citealt{Martin+05}). Finally, to plot
the results in terms of stellar mass we have adopted two linear
relation to convert the \ML s into stellar mass. As test relations
we have recovered from our results at $R_{1}$ and $R_{2}$ the
following best relations: $\log \ML = -0.75 + 0.11 \log \mst$ and
$\log \ML = -1.08 + 0.15 \log \mst$. Moreover, the best fitted
relation between stellar mass and \Ngr\ have been derived and used
to plot the shift in the gradients in term of \Ngr.

We define the quantity $\nabla_{X,OUT}-\nabla_{X,IN}$, where $X=
(age, Z)$, IN and OUT are the input and output X values. In Fig.
\ref{fig: fig_app_1} we show the $\nabla_{X,OUT}-\nabla_{X,IN}$ as
a function of \mst\ and \Ngr, for both age and metallicity and the
two \ML-\mst\ relations. We find that, adding the NIR or UV data,
the gradients, independently from the shift $\delta$ adopted, are
perfectly recovered with a very little scatter. With optical data
only, the uncertainties are larger and some systematic shifts
emerge, mainly for $\delta \geq 0.03$. In this case, our predicted
\gage\ and \gZ\ are underestimated and overestimated with respect
to the input values, respectively. These discrepancies are larger
for galaxies with larger \mst\ and \Ngr, and in the worst case can
amount to $\sim 0.1$ with respect to the ones with lower \mst\ and
\Ngr, for $\delta = 0.05$. Although these findings would led to a
possible flattening of the trends of \gZ\ in terms of mass and
\Ngr, a weaker trend would survive. If one assumes a constant
relation between stellar mass and \Ngr, as for satellites, the
trend with \Ngr\ in Fig. \ref{fig: fig_app_1} would be canceled
and a simple offset in gradient trends would arise in Fig.
\ref{fig: fig2}.

\end{document}